\documentclass[12pt]{article}
\usepackage{epsfig}
\usepackage{amssymb}
\catcode`\@=11
\def\@normalsize{\@setsize\normalsize{15pt}\xiipt\@xiipt
\abovedisplayskip 14pt plus3pt minus3pt%
\belowdisplayskip \abovedisplayskip
\abovedisplayshortskip  \z@ plus3pt%
\belowdisplayshortskip  7pt plus3.5pt minus0pt}
\def\small{\@setsize\small{13.6pt}\xipt\@xipt
\abovedisplayskip 13pt plus3pt minus3pt%
\belowdisplayskip \abovedisplayskip
\abovedisplayshortskip  \z@ plus3pt%
\belowdisplayshortskip  7pt plus3.5pt minus0pt
\def\@listi{\parsep 4.5pt plus 2pt minus 1pt
            \itemsep \parsep
            \topsep 9pt plus 3pt minus 3pt}}
\def\underline#1{\relax\ifmmode\@@underline#1\else
        $\@@underline{\hbox{#1}}$\relax\fi}
\@twosidetrue \relax \catcode`@=12
\catcode`\@=11

\def\section{\@startsection{section}{1}{\z@}{3.5ex plus 1ex minus
   .2ex}{2.3ex plus .2ex}{\large\bf}}

\def\ps@headings{\def\@oddfoot{}\def\@evenfoot{}
\def\@oddhead{\hbox{}\hfill
        \makebox[.5\textwidth]{\raggedright\ignorespaces --\thepage{}--
        \hfill }}
\def\@evenhead{\@oddhead}
\def\subsectionmark##1{\markboth{##1}{}}
} \ps@headings \catcode`\@=12 \relax

\def\figcap{\section*{Figure Captions\markboth
        {FIGURECAPTIONS}{FIGURECAPTIONS}}\list
        {Fig. \arabic{enumi}:\hfill}{\settowidth\labelwidth{Fig. 999:}
        \leftmargin\labelwidth
        \advance\leftmargin\labelsep\usecounter{enumi}}}
 \relax
\def\tablecap{\section*{Table Captions\markboth
        {TABLECAPTIONS}{TABLECAPTIONS}}\list
        {Table \arabic{enumi}:\hfill}{\settowidth\labelwidth{Table 999:}
        \leftmargin\labelwidth
        \advance\leftmargin\labelsep\usecounter{enumi}}}
 \relax
\def\reflist{\section*{References\markboth
        {REFLIST}{REFLIST}}\list
        {[\arabic{enumi}]\hfill}{\settowidth\labelwidth{[999]}
        \leftmargin\labelwidth
        \advance\leftmargin\labelsep\usecounter{enumi}}}
 \relax
\catcode`\@=11
\def\marginnote#1{}
\newcount\hour
\newcount\minute
\newtoks\amorpm
\hour=\time\divide\hour by60 \minute=\time{\multiply\hour by60
\global\advance\minute by- \hour}
\edef\standardtime{{\ifnum\hour<12 \global\amorpm={am}%
    \else\global\amorpm={pm}\advance\hour by-12 \fi
    \ifnum\hour=0 \hour=12 \fi
    \number\hour:\ifnum\minute<100\fi\number\minute\the\amorpm}}
\edef\militarytime{\number\hour:\ifnum\minute<100\fi\number\minute}
\def\draftlabel#1{{\@bsphack\if@filesw {\let\thepage\relax
  \xdef\@gtempa{\write\@auxout{\string
    \newlabel{#1}{{\@currentlabel}{\thepage}}}}}\@gtempa
    \if@nobreak \ifvmode\nobreak\fi\fi\fi\@esphack}
     \gdef\@eqnlabel{#1}}
\def\@eqnlabel{}
\def\@vacuum{}
\def\draftmarginnote#1{\marginpar{\raggedright\scriptsize\tt#1}}
\def\draft{\oddsidemargin -.5truein
        \def\@oddfoot{\sl preliminary draft \hfil
        \rm\thepage\hfil\sl\today\quad\militarytime}
        \let\@evenfoot\@oddfoot \overfullrule 3pt
        \let\label=\draftlabel
        \let\marginnote=\draftmarginnote
\def\@eqnnum{(\theequation)\rlap{\kern\marginparsep\tt\@eqnlabel}%
\global\let\@eqnlabel\@vacuum}  }
\def\preprint{\twocolumn\sloppy\flushbottom\parindent 1em
        \leftmargini 2em\leftmarginv .5em\leftmarginvi .5em
        \oddsidemargin -.5in    \evensidemargin -.5in
        \columnsep 15mm \footheight 0pt
        \textwidth 250mmin      \topmargin  -.4in
        \headheight 12pt \topskip .4in
        \textheight 175mm
        \footskip 0pt
\def\@oddhead{\thepage\hfil\addtocounter{page}{1}\thepage}
        \let\@evenhead\@oddhead \def\@oddfoot{} \def\@evenfoot{}
}
\def\titlepage{\@restonecolfalse\if@twocolumn\@restonecoltrue\onecolumn
     \else \newpage \fi \thispagestyle{empty}\c@page\z@
        \def\thefootnote{\fnsymbol{footnote}} }
\def\endtitlepage{\if@restonecol\twocolumn \else  \fi
        \def\thefootnote{\arabic{footnote}}
        \setcounter{footnote}{0}}  
\catcode`@=12 \relax
\def\ps@headings{\def\@oddfoot{}\def\@evenfoot{}
\def\@oddhead{\hbox{}\hfill
        \makebox[.5\textwidth]{\raggedright\ignorespaces --\thepage{}--
        \hfill }}
\def\@evenhead{\@oddhead}
\def\subsectionmark##1{\markboth{##1}{}}
} \ps@headings \relax
\newcommand{\newc}{\newcommand}
\newc{\ra}{\rightarrow}
\newc{\lra}{\leftrightarrow}

\def\la{\lambda}
\newc{\ome}{\omega}
\newc{\nn}{\nonumber}
\newc{\ba}{\begin{eqnarray}}
 \newc{\ea}{\end{eqnarray}}

\begin{document}
\def\firstpage#1#2#3#4#5#6{
\begin{titlepage}
\nopagebreak
\title{\begin{flushright}
        \vspace*{-0.8in}
\end{flushright}
\vfill {#3}}
\author{\large #4 \\[1.0cm] #5}
\maketitle \vskip -7mm \nopagebreak
\begin{abstract}
{\noindent #6}
\end{abstract}
\vfill
\begin{flushleft}

\end{flushleft}
\thispagestyle{empty}
\end{titlepage}}

\def\simlt{\stackrel{<}{{}_\sim}}
\def\simgt{\stackrel{>}{{}_\sim}}
\date{}
\firstpage{3118}{IC/95/34} {\large\bf Instanton induced charged fermion and
neutrino masses in a minimal Standard Model scenario from intersecting  D-branes}
 {G. K. Leontaris}
{\normalsize\sl Theoretical Physics Division, Ioannina
University,
GR-45110 Ioannina, Greece\\
\\ [2.5mm]
 }
{String instanton Yukawa corrections from  Euclidean D-branes are investigated in an
effective Standard Model theory obtained from the minimal $U(3)\times U(2)\times U(1)$
D-brane configuration. In the case of the minimal chiral and Higgs spectrum,  it is found
that superpotential contributions are induced by string instantons for the perturbatively 
forbidden entries of the up and down quark mass matrices.  Analogous non-perturbative 
effects  generate  heavy Majorana neutrino  masses and a Dirac neutrino 
texture with  factorizable Yukawa  couplings. For this latter case, a specific example 
is worked out where it is shown how this texture can reconcile the neutrino data.
 }

\vskip 3truecm

\newpage

\section{Introduction}

The recent years  D-brane configurations with $U(3)\times U(2)\times U(1)^n$ gauge symmetry have been
systematically  investigated and extensively analyzed~\cite{Antoniadis:2000ena}-\cite{Anastasopoulos:2006da},
particularly in  the  context of intersecting D-branes. This gauge symmetry could be considered as the natural
successor of the Standard Model (SM) gauge group $G_{SM}=SU(3)_C\times SU(2)_L\times U(1)_Y$ in the context of
 D-branes.   The constant endeavors to examine the low energy implications of the emerging effective low energy
 theory  during the  last few years have been absolutely justified, as many of these efforts were crowned with
 considerable progress towards a successful D-brane realization of the SM gauge group with its chiral
 and Higgs matter content.

In the intersecting D-brane constructions the SM chiral matter arises in the intersection of the brane-stacks
 wrapping three cycles in the internal six-dimensional manifold.  Wrapping branes on a compact manifold
intersect each other, while the  number of fermion fields  equals the number of intersections, a fact that
eventually reveals a possible geometrical origin of the fermion family multiplicity. The emerging effective
field theory model usually consists of the SM gauge symmetry augmented by some global $U(1)$ symmetries,
with the minimal fermion and Higgs spectrum escorted (in several cases) by some  additional -hopefully
innocuous- matter representations.

One of the main shortcomings of these  constructions is the absence of several Yukawa
couplings necessary to provide all quark and lepton fields with non-zero masses. Indeed,
a generic feature in D-brane constructions is that  ordinary matter fields are charged
under the aforementioned  global $U(1)$ symmetries accompanying the SM gauge group. Undoubtedly,
these  symmetries impose additional restrictions (beyond those of the SM gauge symmetry) on the
 perturbative  Yukawa superpotential  terms and as a consequence, some of the SM fermion fields
remain massless.

It was shown that these global $U(1)$'s are broken from stringy  instantons effects
\cite{Blumenhagen:2006xt,Ibanez:2006da,Florea:2006si,Abel:2006yk} and as a consequence
new, non-perturbative contributions can be generated to the superpotential. In type IIA
compactifications in particular, the eventual candidates are the Euclidean D2 instantons
(subsequently called ${\cal E}2$ instantons for short) wrapping three cycles in the internal manifold.
If they are chosen to intersect with the relevant D6-branes appropriate number of times,
there exist inextricable superpotential terms filling in the zeros of the fermion mass
matrices. Compared to the perturbative Yukawa couplings,  the instanton induced terms appear
to be  suppressed by the exponential of the classical instanton action $e^{-S_{{\cal E}2}}$.
These corrections although subdominant with respect to the tree-level perturbative
Yukawa couplings, might prove of vital importance as far as the viability of the construction
is concerned.   Indeed, in the class of models under consideration, these instanton induced
terms could  offer a natural explanation for the observed  hierarchy of the fermion mass
spectrum,  acting as a surrogate for missing non-renormalizable terms of the perturbative sector.

In the present work we will examine the stringy instanton effects in the case
of a minimal D-brane set-up required to embed the SM gauge symmetry.
This configuration consists  of three brane-stacks leading to
$U(3)\times U(2)\times U(1)$ gauge symmetry with the ordinary quarks and lepton fields living
in their intersections. It was shown that three different fermion embeddings are
possible\cite{Antoniadis:2004dt}, while, due to the minimal structure of this configuration
some particular cases can be naturally embedded in  higher unified symmetries such as the
 $U(5)$~\cite{Antoniadis:2007jq}, the trinification~\cite{Leontaris:2008mm}
 or the Pati-Salam\cite{Pati:1974yy} D-brane analogue\cite{Leontaris:2000hh}.

The paper is organized as follows. In the next section we briefly present
the derivation of the model from the particular D-brane set-up and  give a short description
of  its embedding in the intersecting D-brane scenario.  We introduce the chiral and Higgs spectrum
and emphasize the salient features of the model, including the unification prospects and the
 perturbative superpotential structure for several choices of the matter content. In section 3,
 we  recapitulate
the basic procedure on the derivation~\cite{Blumenhagen:2006xt,Ibanez:2006da} of the stringy
instanton non-perturbative Yukawa couplings and apply the results to the specific D-brane
construction analyzed in this work. We derive the non-perturbative contributions for the
quark and neutrino mass matrices and comment on the peculiar nature of the resulting Yukawa 
textures. A separate discussion is devoted to the attractive Dirac neutrino mass matrix 
texture in section 4 and emphasis is given in a specific example which reconciles
 the experimental data. Finally, in section 5 we present our conclusions.

\section{The Standard Model on a $U(3)\times U(2)\times U(1)$ D-brane set-up}

We start this section with a brief description of the simplest model constructed
by the minimum possible number of brane-stacks. The SM embedding in a D-brane
configuration has received much attention the last decade and attempts to reproduce
its spectrum and low energy successful reconciliation with all the known experimental
data  has stimulated a thorough investigation of the configurations with several sets
of brane-stacks leading to the enhanced SM symmetry
\ba
U(3)\times U(2)\times U(1)^n\ra SU(3)\times SU(2)\times U(1)^{n+2}\supset G_{SM}
\label{ESM}
\ea
The integer power $n$ represents the number of $U(1)$ branes needed to reproduce the desired
SM spectrum while a linear combination of the $n+2$ abelian factors  will accommodate
the hypercharge generator.\footnote{
From (\ref{ESM}) we can observe that due to the group relation $U(N)\ra SU(N)\times U(1)$,
one might assume that the SM gauge symmetry could be generated with only the two brane stacks
$U(3), U(2)$, so we could take $n=0$. It can be easily shown that the hypercharge generator
cannot be expressed only in terms of these two $U(1)$ symmetries, so we need $n\ge 1$. }
 SM matter fields are `incarnated' through open strings
with their endpoints on the intersecting D6-branes and are localized in the intersection
locus of the latter. In type IIA orientifolds the following types of representations arise:

Open strings appearing in the intersection locus  of two $D6_a,D6_b$ branes
($ab$ for short) `create' bifundamental representations $(N_a,\bar N_b)$ while
their multiplicity is  given by the number of intersections
\ba
I_{ab}&=&\prod_{i=1}^3\left(m_{ai} n_{bi}-m_{bi} n_{ai}\right)\label{ABb}
\ea
The six-dimensional internal manifold is assumed to be factorizable with a torus geometry,
$T^6=T^2_1\times T^2_2\times T^2_3$ while here the $(n_{ai},m_{ai})$ pairs represent the
winding numbers of the $D6_a$-stack around the two radii of the $i^{\rm th}$ torus.

 Chiral fermions appear also  in $(N_a,N_b)$ representations from the intersections of a
 $D6_a$ brane-stack with a mirror brane $\Omega {\cal R}D6_b\ra D6_{b^*}$.  Their number
 is given by
\ba
I_{ab^*}&=&-\prod_{i=1}^3\left(m_{ai} n_{bi}+m_{bi} n_{ai}\right)\label{AB}
\ea
States arising in the intersection of a $D6_a$-brane with its corresponding mirror one,
may belong to the antisymmetric or symmetric representations. Those which remain
 invariant under the combined $\Omega {\cal R}$ action belong to the antisymmetric
 representations of the $U(N_a)$ gauge group and their multiplicity (denoted as
$I_{aa^*}^A$) is given by
\ba
I_{aa^*}^A&=&8m_{a1}m_{a2}m_{a3}\label{Asym}
\ea
Depending on the specific winding numbers, there may also appear symmetric
and antisymmetric representations with equal multiplicities
\ba
{I'}^{A,S}_{aa^*}&=&4m_{a1}m_{a2}m_{a3}\left( n_{a1}n_{a2} n_{a3}-1\right)\label{Sym}
\ea
Finally, additional  restrictions are imposed  on the $(n_{ai}, m_{ai})$ sets
originating from the $RR$-type tadpole conditions which read:
\ba
T_0\;=\;\sum_{a}N_a\,n_{a1}n_{a2}n_{a3}=16 &,&
T_i\;=\;\sum_{a}N_a\,n_{ai}m_{aj}m_{ak}=0,\;\;\;i\ne j\ne k\ne
i\label{TC}
\ea
where the indices $i,j,k$ take the values $1,2,3$ and refer  to the three torii $T^2_j$.

\begin{figure}[!t]
\centering
\includegraphics[width=0.28\textwidth]{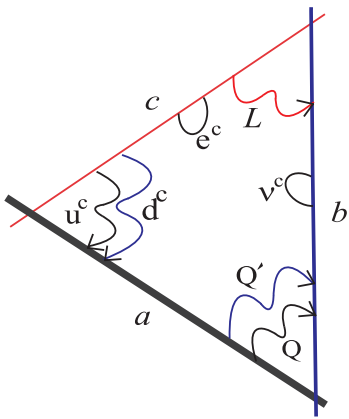}
\caption{\label{fconf}A depiction of the  $U(3)\times U(2)\times U(1)$
D-brane configuration with strings representing the SM states. For the sake of simplicity,
 $a,b,c$ denote  $U(3), U(2)$ and $U(1)$  branes respectively.  In this figure D-branes are not
distinguished from their corresponding mirrors. Thus, the blue string
representing the quark doublet $Q'$ is stretched  between the $D6_a$ and the mirror
 $D6_{b^*}$. Similarly,  one endpoint of the ``$d^c$-string''
 is attached on the mirror $D6_{c^*}$.} \label{f1}
\end{figure}

Returning to (\ref{ESM}), we have already asserted that the  three SM gauge factors need
at least three brane-stacks, however, if one insists to accommodate the SM fermions only
 in bifundamentals, then one should take at least $n=2$~\cite{Anastasopoulos:2006da,Gioutsos:2006fv}
. If, for the sake of simplicity
 and  `economy' one uses the possible minimal brane set-up,  i.e. when $n=1$, some of the
 fermions are accommodated in bifundamentals obtained in the intersections of two different
brane-stacks, while others are  accommodated in antisymmetric or symmetric representations which
can be created by strings with  endpoints on  a given brane-stack and its corresponding
mirror.  As mentioned above, there are three distinct arrangements of the SM-fields
(see \cite{Antoniadis:2004dt}, \cite{Gioutsos:2006fv}) and here we concentrate on the
configuration depicted in figure \ref{f1}.
For later convenience we call $a,b,c$ the stacks related to $U(3), U(2),U(1)$, respectively. We
start exploring the model implementing the minimal fermion and Higgs spectrum of our particular
D-brane set-up which is shown in Table \ref{U321}.
This consists of the three SM fermion generations, their corresponding right handed neutrinos
and one pair of Higgs doublets.
\begin{table}[!t]
\centering
\renewcommand{\arraystretch}{1.2}
\begin{tabular}{ccrrrr}
\hline
Inters. & $SU(3)\times SU(2)$& ${\cal Q}_a$ & ${\cal Q}_b$ & ${\cal Q}_c$ &$Y$\\
\hline
 $ab$   & $1\times Q\,(3,\bar 2)$ & $1$ & $-1$ & $0$ &$\frac 16$   \\
 $ ab^*$    & $2\times Q'\,(3, 2)$ & $1$ & $1$ & $0$& $\frac 16$   \\
$ac$& $3\times u^c\,(\bar 3,1)$ & $-1$ & $0$        & $1$&$-\frac 23$\\
$ ac^*$ &$3\times d^c\,(\bar 3,1)$ & $-1$ & $0$ & $-1$ &$\frac 13$\\
$bc$& $3\times L\,(1,\bar 2)$ & $0$ & $-1$ & $1$   &$-\frac 12$     \\
 $cc^*$   & $3\times e^c\,(1,1)$ & $0$ & $0$ & $-2$    &$1$    \\
 $bb^*$    & $3\times \nu^c\,(1,1)$ & $0$ & $-2$ & $0$  &$0$      \\
\hline
 $bc^*$   &      $\left.
                  \begin{array}{r}1\times H_d(1,2)\\
                                  1\times {H}_u(1,\bar 2)
                 \end{array}\right.$
          &      $\left.
                 \begin{array}{r}
                 0 \\
                 0
                 \end{array}\right.$
         &       $\left.
                 \begin{array}{r}1\\
                 -1
                 \end{array}\right.$
         &       $\left.\begin{array}{r}
                 1\\
                 -1
                 \end{array}\right.$
         &       $\left.
                 \begin{array}{r}
                 -\frac 12  \\
                  \frac 12
                  \end{array}
                  \right.$\\
\hline
\end{tabular}
 \caption{The quantum numbers of the SM fermions in the $U(3)\times
U(2)\times U(1)$ brane configuration. The last column is the Hypercharge
$Y=\frac 16{\cal Q}_a-\frac 12{\cal Q}_c$ while three  previous
ones refer to the $U(1)$ charges with respect to the $a,b,c$ brane-stacks.\label{U321} }
\end{table}
The spectrum is consistent with the following hypercharge definition
\ba
Y&=&\frac 16 {\cal Q}_a-\frac 12\,{\cal Q}_c
\ea
where  ${\cal Q}_{a,c}$ are the  $U(1)_{a,c}$ charges of the SM representations
while the corresponding $U(1)_Y$ gauge coupling $a_Y$ is the combination
\ba
 \frac{1}{a_Y}&=&\frac{1}{6}
\frac{1}{a_3}+\frac 12
\frac{1}{a_1} \label{gY}
\ea
It turns out that the U(1)-normalization coefficient is given by
$k_Y=\frac{1}{6}\xi+\frac 12\xi'$
where
$\xi=\frac{\alpha_2}{\alpha_3},\xi'=\frac{\alpha_2}{\alpha_1}$
are the gauge coupling ratios from which we deduce  the Weinberg angle at the String scale
$M_S$,
$$\sin^2\theta_W=\frac{1}{1+k_Y}=\frac{6}{6+\xi+3\xi'}$$
In the particular case of a unified value of all three gauge couplings $a_1,a_2,a_3$,
the string scale $M_S$ turns out to be rather high. For partial unification $a_2=a_3$
the `standard' GUT value $\sin^2\theta_W=\frac 38$ could be obtained for $\xi'=3$, whilst
for the standard MSSM spectrum and a SUSY scale $m_s\sim m_Z$, non-abelian gauge
unification ($a_2=a_3$) implies
\ba
M_S&=&e^{\frac{\pi}2\left(\frac{\sin^2\theta_W}{\alpha}-\frac{1}{\alpha_3}\right)}
\left(\frac{m_Z}{m_s}\right)^{\frac{13}{12}} m_Z\sim \,{\cal O}(10^{16}){\rm GeV}
\ea
In D-brane scenarios each gauge group factor is related to a
brane-stack which occupies certain dimensions of the higher
dimensional space. Since the present model is based on a product of gauge
groups, these correspond to an equivalent number of brane-stacks
which in general span different directions of the 10-d space.  Thus,
gauge couplings in general have different values and the String scale may be
significantly lower in intersecting scenarios.
A detailed analysis of the unification prospects and many other low energy
phenomenological issues in the context of intersecting branes has been presented
elsewhere~\cite{Leontaris:2007ej}.

In the present setup, one quark doublet  (namely $Q=(3,\bar 2)$), the right-handed up-quarks $u^c$,
and the lepton doublets ${\cal L}$ arise from the intersection of $D6_a-D6_b$, $D6_a-D6_c$ and
 $D6_b-D6_c$ branes respectively. In this case, their number is equal to
the corresponding intersections $I_{ab}$, $I_{bc}$, $I_{ac}$, determined by the formula
(\ref{ABb}).

Two quark doublets belong to the $(3,2)$ representation of the $U(3)\times U(2)$ brane-stacks thus
 they arise from the intersections of $D6_a$ with the mirror D-brane $\Omega {\cal R}D6_b\ra D6_{b^*}$,
hence their number is determined by  the formula (\ref{AB}). The remaining chiral states arise in the intersection
of a  $D6_x$-brane ($x=b,c$) with its corresponding mirror one, while they could  belong to the antisymmetric or
symmetric representations, their multiplicity denoted $I_{xx^*}^A$ and $I_{xx^*}^S$ respectively and given by
(\ref{Asym}) and (\ref{Sym}) analogously.
In particular,  right-handed electrons arise in the $c{c^*}$ intersection and they belong to the antisymmetric
representation, their $U(1)_c$ charge is  ${\cal Q}_{c}=-2$, its absolute value being twice as big compared to
that of a bifundamental. In the  $bb^*$ intersection on the other hand, since the $U(1)_b$ generator does not
participate in the hypercharge, the corresponding antisymmetric representations are electrically neutral and
can be identified with the right handed neutrinos, while their $U(1)_b$-charge can have any of the two possible
values ${\cal Q}_{b}=\pm 2$.

As can be inferred from the above discussion, we have chosen in this minimal approach to derive the quark doublets
from two different intersections. Thus, in Table~\ref{U321} we observe  that one left-handed quark doublet is in
fact an $SU(2)$-antidoublet since it arises from the $ab$ sector, while the remaining two quark families are
$SU(2)$-doublets, since they come from the $ab^*$ sector. The reason for this arrangement~\cite{Ibanez:2001nd}
is that the $SU(N)$ cancelation conditions demand the same number of $N$ and $\bar N$'s for any $U(N)$ gauge
 group factor  and this has to be satisfied even for the $SU(2)$ gauge group, hence one has to distinguish
 between doublets and anti-doublets so we can ensure that they appear in  equal numbers. Thus, a consistent
derivation of the minimum SM spectrum of the model requires $I_{ab}=1, I_{ab^*}=2$ for the left-handed quarks,
 $I_{ac}=I_{ac^*}=-3$ for up and down right handed quarks, and $I_{bc}=-3$ for the lepton doublets.
Furthermore, the conditions $I_{cc^*}=I_{bb^*}=-3$ ensure the existence of three right-handed electrons $e^c$
 and neutrinos $\nu^c$ and finally $I_{bc^*}^{N=2}=1$ gives one  MSSM Higgs pair\footnote{We should note that
 the last three conditions, and in particular $I_{cc^*}=I_{bb^*}=-3$ are not mandatory and can be replaced
 with the less tight conditions $|I_{cc^*}|, |I_{bb^*}|>3$. The existence of an extra right-handed electron
  ${e'}^c$ and a neutrino ${\nu'}^c$ , under certain circumstances could explain the reported high energy
 positron excess through a coupling $e^c\, {e'}^{c*}{\nu'}^c$\cite{Adriani:2008zr}.}.
The wrapping numbers $(n_{ai},m_{ai})$  are also subject to  additional restrictions  from Tadpole cancelation
 conditions (\ref{TC}) which usually impose restrictions on extra hidden matter related to strings attached to
 bulk branes.

The tree-level quark and lepton Yukawa couplings of this minimal chiral and Higgs spectrum which is consistent
with the above hypercharge assignment are
 \ba
 {\cal W}\supset \lambda^u_{pj}\,Q_p'\,u^c_j\, H_u
 +(\lambda^d_j\,Q\,d^c_j+\lambda^l_{ij}\,L_i\,e^c_j)H_d
\label{SP}
 \ea
 Let us state a few remarks for the resulting superpotential, starting from the quark sector.
  In the first term, the indices $i,j$ run over all three fermion generations, while $p$  takes
 only two values. Thus, only two doublet quark flavors contribute through tree-level perturbative
 Yukawa couplings in the up-quark mass matrix. The reason  is that the additional $U(1)_{(a,b,c)}$
 charges carried by the various representations do not allow for a coupling involving the
 representation $Q(3,\bar 2)_{(1,-1,0)}$ generated in the intersection of the $ab$
 brane-stacks. For the same reason tree-level mass terms for the two quark doublets living
 in the intersection $ac^*$ do not appear in the down quark mass matrix, since  the down
 right-handed quarks couple solely to the remaining quark doublet.

  Depending on the particular  assignment of the fermion generation, three possible mass matrix
   textures arise. Thus, if the  two $Q'_p$ quark doublets correspond to the first two generations,
   then we obtain the following up and down quark Yukawa textures
\ba
m_{Q}=\left(
\begin{array}{lll}
 \la_{11}^u &\la_{12}^u & \la_{13}^u \\
  \la_{21}^u &\la_{22}^u & \la_{23}^u
   \\
0 & 0&0
\end{array}
\right)\langle H_u\rangle,&&m_{d}=\left(
\begin{array}{lll}
0&0&0\\
 0 &0 & 0 \\
  \la_{31}^d &\la_{32}^d & \la_{33}^d
\end{array}
\right)\langle H_d\rangle\label{ca1}
\ea
 Clearly, in the same way, we have the following two remaining distinct possibilities, namely
\ba
m_{Q}=\left(
\begin{array}{lll}
 \la_{11}^u &\la_{12}^u & \la_{13}^u \\
0 & 0&0 \\
  \la_{31}^u &\la_{32}^u & \la_{33}^u
\end{array}
\right)\langle H_u\rangle,&&m_{d}=\left(
\begin{array}{lll}
0&0&0\\
  \la_{21}^d &\la_{22}^d & \la_{23}^d\\
 0 &0 & 0
\end{array}
\right)\langle H_d\rangle\label{ca2}
\ea
and
\ba
m_{Q}=\left(
\begin{array}{lll}
0 & 0&0 \\
 \la_{21}^u &\la_{22}^u & \la_{23}^u \\
  \la_{31}^u &\la_{32}^u & \la_{33}^u
\end{array}
\right)\langle H_u\rangle,&&m_{d}=\left(
\begin{array}{lll}
  \la_{11}^d &\la_{12}^d & \la_{13}^d\\
  0&0&0\\
 0 &0 & 0
\end{array}
\right)\langle H_d\rangle\label{ca3}
\ea
As can be seen, there is a complementary texture zero structure of the up and down quark
mass matrices at the perturbative level, in the sense that the zero entries of the first are
non-zero in the second and vice versa. The correct choice of family assignment will be dictated
when the  order of magnitude of the instanton and/or possible other contributions will be specified.
Thus, a conclusive answer to this question, depends on the particular mechanism which will be adopted
 to generate the missing terms.

As far as the lepton sector is concerned,  Yukawa terms exist for all charged leptons, but
 Dirac and Majorana mass terms for the neutrinos are absent, hence they remain massless at
tree level.

There are several  possible solutions to this issue, among them the most obvious are those
involving suitable extensions of the Higgs sector~\footnote{
Of course we may adopt a more radical approach and assume that the $SU(2)$  cancelation
conditions are satisfied by an appropriate number of $SU(2)$ antidoublets  living in  the
Hidden sector\cite{Ibanez:2008my}. Then,  all left-handed quarks could  originate from the
same sector in the intersection of the $D6_a, D6_b$ brane-stacks, so that  tree-level
couplings for all families are permitted.}.
The most exciting and economic alternative however, is that of invoking stringy
instanton effects to fill in the gaps in the mass matrices and create other useful couplings.

 We start the analysis with the first alternative. Our initial suggestion was confined to the
 D-brane realization of the SM model  with exactly the minimal chiral content  and only
 one pair of Higgs doublets. As a result, a number of useful Yukawa couplings were
 missing at the perturbative superpotential. The most obvious possibility of deriving
 the requested couplings is to extend  the Higgs sector introducing new multiplets with
 suitable  $U(1)$ charges.
The additional fields with their corresponding quantum numbers are shown in Table \ref{T2}.
These include an additional right-handed singlet neutrino supermultiplet with $U(1)_b$ charge
${\cal Q}_b=+2$ and one more  Higgs pair emerging in the intersection of branes $D6_b$ and $D6_c$.
 \begin{table}[!t]
\centering
\renewcommand{\arraystretch}{1.2}
\begin{tabular}{ccrrrr}
\hline
Inters. & $SU(3)\times SU(2)$& ${\cal Q}_a$ & ${\cal Q}_b$ & ${\cal Q}_c$&$Y$ \\
\hline
 $bb^*$    & $ {\nu'}^c\,(1,1)$ & $0$ & $+2$ & $0$   &$0$     \\
\hline
 $bc$   & $\left.\begin{array}{r} H'_d(1,2)\\
                 H'_u(1,\bar 2)
                 \end{array}\right.$  & $\left.\begin{array}{r}0\\
                 0
                 \end{array}\right.$ &$\left.\begin{array}{r}-1\\
                 1
                 \end{array}\right.$& $\left.\begin{array}{r}1\\
                 -1
                 \end{array}\right.$
                 &       $\left.
                 \begin{array}{r}
                 -\frac 12  \\
                  \frac 12
                  \end{array}
                  \right.$ \\
\hline
\end{tabular}
 \caption{The additional Higgs representations with their quantum numbers. \label{T2} }
\end{table}
Taking into account the fields of Table \ref{T2}, the following tree-level Yukawa terms should
be included to the superpotential
\ba
{\cal W}'&=&{\lambda'}_j^u\,Q\,u_j^c\,H_u'+{\lambda'}_j^d\,Q'_p\,d^c_j\,H_d'
+{\lambda'}_{ij}^{\nu}\,{\cal L}_i\,{\nu'}^c_jH_u
\label{SPprime}
\ea
These new terms are sufficient to provide the missing entries in the quark mass matrices and
generate the Dirac-type neutrino masses.

It is worth mentioning that the introduction of the second neutrino supermultiplet allows
for an alternative option. It can be seen that the following fourth order non-renormalizable
terms are permitted by the SM gauge and the three global $U(1)$ symmetries
\ba
\frac{1}{M_S}Q\,u_j^c\,H_u\,{\nu'}^c+\frac{1}{M_S}\,Q'_p\,d^c_j\,H_d\,\nu^c\label{newY}
\ea
Assuming a vacuum expectation value to a combination $\tilde \nu_H$ of the scalar partners
$(\tilde\nu^c,\tilde{\nu'}^c)$ we observe that
both terms are contributing to the zero entries of the quark mass matrices. In this
case one dispenses with the use of the second Higgs pair, since in the new couplings (\ref{newY})
only the first pair of Higgs fields participates. This  second choice however, discriminates between the
tree-level entries (\ref{SP}) and the new contributions (\ref{newY}) in the mass matrices, since
the latter are suppressed relative to the first by the factor $\frac{\langle {\tilde\nu}_H\rangle}{M_S}$
and considerable adjustment of the parameters is needed to derive the observed quark mass hierarchy.

Next, instead of expanding the chiral and Higgs spectrum, as a second alternative, we consider
 stringy instantons  and  take into account  their induced non-perturbative contributions
 to the perturbative superpotential (\ref{SP}).  As in the case of the non-renormalizable
 terms (\ref{newY}), these corrections are also expected to be suppressed relative to
 the  tree-level entries, the suppression factor being now proportional to the exponential
 of the volume of the cycle wrapped by the instanton in the internal manifold.
  We analyze these issues in the next section.

\section{String Instanton induced masses}

We have already realized that the D-brane embedding of the SM implies additional  restrictions
on the Yukawa superpotential due to the presence of the three $U(1)_{a,b,c}$ abelian symmetries.
As a consequence, a number of useful Yukawa couplings crucial for  the fermion mass terms are
absent at the tree level of the perturbative sector.

In general, if $\Phi_{a_jb_j}, j=1,2,\dots, J$ represent  superfields generated
in the intersection of  $D6$ brane-stacks, it is possible that the coupling
$\prod_j\Phi_{a_jb_j}$ although invariant with respect to the non-abelian part
of the gauge symmetry, might violate  some abelian global $U(1)_a$ factor.

It has been recently suggested~\cite{Blumenhagen:2006xt,Ibanez:2006da} that in viable models,
the missing tree-level Yukawa couplings could be generated from non-perturbative effects.
In particular, considering ${\cal E}2$ instantons in type IIA string theory having appropriate
number of intersections with the $D6$-branes, a non-perturbative term
\ba
{\cal W}_{n.p.}&\supset&\prod_{j=1}^J\Phi_{a_jb_j}\,e^{-S_{\cal E}}\label{instcoup}
\ea
is induced, where the instanton action $S_{\cal E}$ can absorb the $U(1)_a$ charge
excess of the matter fields operator $\prod_j\Phi_{a_jb_j}$. The global $U(1)_a$ charges
carried by the instanton action have their microscopic origin in the extra fermionic zero
 modes living in the intersection of the ${\cal E}2$-brane with the $D6_a$ brane-stack.
 Under the aforementioned abelian symmetry the transformation property of the exponential
instanton action is
\ba
\,e^{-S_{\cal E}}&\ra&\,e^{-S_{\cal E}}\,e^{\imath\,{\cal Q}_a({\cal E}2)\Lambda_a}
\ea
where ${\cal Q}_a({\cal E}2)$ represents the amount of the $U(1)_a$-charge violation
by the ${\cal E}2$ instanton. For an instanton wrapping appropriate number of
cycles, this can be made exactly opposite to the total charge of
the operator $\prod_j\Phi_{a_jb_j}$, so the coupling (\ref{instcoup}) is also
$U(1)_a$-invariant. Besides, the induced coupling (\ref{instcoup}) involves an
exponential suppression by the classical instanton action
${\cal W}_{n.p.}\propto \exp\{-\frac{8\pi^2{\rm Vol}_{\cal E}}{g_a^2{\rm Vol}_{D6_a}}\}$ which, as can  be
seen depends on the volume Vol${}_{{\cal E}2}$ of the cycle wrapped by the instanton and
is inversely proportional to the perturbative string gauge coupling $g_a^2$. It is
 expected  that even in the case of small $g_a$, string instanton contributions
might become relevant.

In  computing  these non-perturbative effects the important ingredients are the instanton
 fermionic zero modes. We distinguish two different classes of them with respect
 to the charge they carry under the abelian gauge factors. In one class
  there are two uncharged fermionic zero modes $\theta_i$ with both endpoints
  attached on the ${\cal E}2$ instanton and they correspond to the two broken
  supersymmetries. In a second class are classified all the charged fermionic zero modes
  which are of  primary importance in the computation of the instanton effects. These appear
  in the intersection of the ${\cal E}2$ instanton with a certain $D6_a$ brane-stack, while their
  relevance to a specific superpotential coupling is related to the number of ${\cal E}2-D6_a$
  intersections. In particular, the crucial quantity which appears in the coupling (\ref{instcoup})
  is the  charge ${\cal Q}_a({\cal E}2)$ which is related to the ${\cal E}2$ intersections with the
$D6$ brane stacks as follows. The ${\cal E}2$ instanton brane is considered wrapping a homological
three-cycle $\pi_{\cal E}$ in the internal manifold, while is localized in the four-dimensional
space-time. If $\pi_a,\pi_{a^*}$ are the homological three-cycles of the $D6_a$ brane-stack and its
mirror respectively, then
\ba
{\cal Q}_a({\cal E}2)&=&-{\cal N}_a\,\pi_{\cal E}\circ (\pi_a-\pi_{a^*})\equiv -{\cal N}_a\,
(I_{{\cal E}a}-I_{{\cal E}{a^*}})\label{instcha}
\ea
where the $I_{{\cal E}a}$ and  $I_{{\cal E}{a^*}}$ stand for the relevant intersection numbers.
In what follows, we will consider a class of rigid $O(1)$ instantons, wrapping a rigid
orientifold-invariant cycle in the internal space, where due to the ${\cal E}2-a$ and
${\cal E}2-a^*$ identification the charge (\ref{instcha}) simplifies to
\ba
{\cal Q}_a({\cal E}2)&=&-{\cal N}_a\,\pi_{\cal E}\circ \pi_a\equiv -{\cal N}_a\,
I_{{\cal E}a}\label{E2D6}
\ea
Assuming appropriate number of wrappings, it is possible to cancel exactly the
$U(1)_a$ charge excess and make the desired coupling invariant.

Returning to the specific model discussed in this work, we can check from
(\ref{SP}) that there are several missing superpotential couplings, including
those  for the up- and down quarks  and the right-handed Majorana
neutrino masses. The absence of these terms make the model rather unrealistic
at the perturbative level. We will see that instanton contributions do play a vital
role on the issues of viability and phenomenological consistency of this construction.

We start with  the up quark coupling $ Qu^c_jH_u$  which, as can be observed,
violates the  $U(1)_b$ charge by two units
$$ {\cal Q}_{b_Q}+{\cal Q}_{b_{u^c}}+{\cal Q}_{b_{H_u}}=-2$$
 and therefore cannot exist at tree-level. Assuming a ${\cal E}2$ instanton
 intersecting with the $D6_b$ brane-stack, so that
\ba
{\cal Q}_b({\cal E}2)&=&
-{\cal N}_b\, I_{{\cal E}b}
\ea
we conclude that  $I_{{\cal E}b}=-1$ eliminates the charge-excess and the relevant coupling
is allowed. The intersections of ${\cal E}2$ with the relevant branes is depicted in fig \ref{f1}.
 At the computational level, one has to integrate over all instanton zero modes, the
 path integral being
\ba
\int\,\{d^4x\,d^2\theta\,d^2\lambda_b\}\,e^{-S_{\cal E}}\,y_j^u
\epsilon_{mn}\epsilon_{r\ell}<\lambda_b^mQ^nu^c_jH_u^r\lambda_{b^*}^{s}>\,e^{Z'}\label{IntQu}
\ea
The integration (\ref{IntQu})  is over the four bosonic $x^{\mu}$  and two fermionic $\theta^i$
zero modes as well as the two instanton modes $\lambda_b$. Further there is an overall exponential
suppression of  the instanton action which makes the coupling invariant and $e^{Z'}$ in the
regularized one-loop amplitude. Finally, $m,n,r,s$ are $SU(2)$ indices, while $y^u_j$ is
a coupling constant with $j$ running over all family indices. Clearly, the above instanton induced
terms contribute to the zeros of the up-quark matrices presented in the previous section.
\begin{figure}[!t]
\centering
\includegraphics[width=0.35\textwidth]{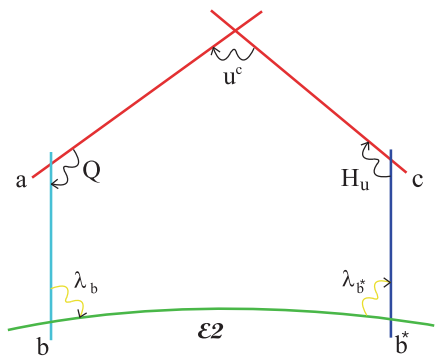}
\caption{\label{fconf}Stringy instanton contributions to the up-quark Yukawa mass matrix.} \label{f2}
\end{figure}

We turn now our attention to the down quark mass matrix. The trilinear couplings $Q'_pd^c_jH_d$
violate the $U(1)_b$ charge by two units ${\cal Q}_{b_{Q'}}+{\cal Q}_{b_{d^c}}+{\cal Q}_{b_{d_u}}=2$,
hence this term is also forbidden in the perturbative superpotential. In an analogous manner with the
up-quark case discussed above, we introduce here a second instanton ${\cal E}2'$ which intersects with
$D6_b$ and $D6_{b^*}$ brane-stacks, so that ${\cal Q}_b({\cal E}2')=
-{\cal N}_b\, I_{{\cal E}2'b}$, with $I_{{\cal E}2'b}=+1$. An integration over
the instanton zero modes as in the case of (\ref{IntQu}) will lead to the effective instanton induced
down quark Yukawa coupling $Q'_pd^c_jH_d$, suppressed by analogous exponential factors.  Summarizing,
the non-perturbative contributions to quark mass matrices induced by string instantons are
\ba
{\cal W}_{n.p.}&=&{\lambda'}^u_{j}Q\,u^c_jH_u+{\lambda'}^d_{pj}Q'_pd^c_jH_d\label{npQ}
\ea
where the couplings ${\lambda'}^u_{j}$ and ${\lambda'}^u_{j}$ are  suppressed
compared to the perturbative ones, due to the
exponential factors involving the classical instanton action. Taking into account the
non-perturbative contributions the mass matrices obtain the following texture form
\ba
m_{Q}=\left(
\begin{array}{lll}
 \la_{11}^u &\la_{12}^u & \la_{13}^u \\
  \la_{21}^u &\la_{22}^u & \la_{23}^u
   \\
  {\la'}_{1}^{u} & {\la'}_{2}^{u} &  {\la'}_{3}^{u}
\end{array}
\right)\langle H_u\rangle,&&m_{d}=\left(
\begin{array}{lll}
 {\la'}_{11}^{d} & {\la'}_{12}^{d} &  {\la'}_{13}^{d}\\
 {\la'}_{21}^{d} & {\la'}_{22}^{d} &  {\la'}_{23}^{d}\\
  \la_{31}^d &\la_{32}^d & \la_{33}^d
\end{array}
\right)\langle H_d\rangle
\ea
and analogously for the other two cases (\ref{ca2},\ref{ca3}). We should point out
that in general,  fermion mass textures as those above are rather unusual and
considerable adjustment is needed to obtain the known up and down quark mass hierarchies.
The rather striking fact however, is that since the non-perturbative couplings
are suppressed by factors of the form $e^{-S_{{\cal E}}}$,
relative to the perturbative ones,  the quark mass ratios are directly
 correlated to the  geometry of the internal manifold.

\section{Neutrino masses}

\begin{figure}[!t]
\centering
\includegraphics[width=0.65\textwidth]{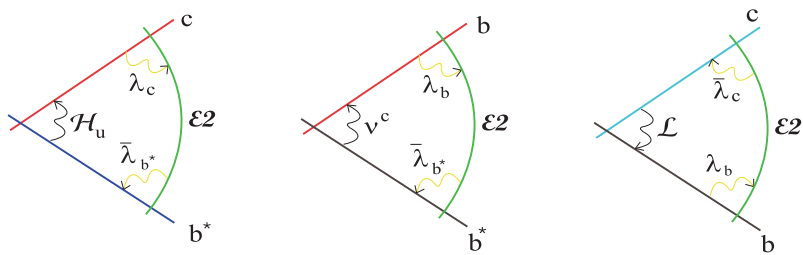}
\caption{\label{fconf}The zero modes on the ${\cal E}2-D6$ intersections generating an effective
Dirac neutrino mass term.} \label{f3}
\end{figure}

 The experimental evidence of non-zero neutrino masses and mixing urges for a rather profound
 extension of the Standard Model, i.e., the introduction of the right-handed neutrino. This
 makes possible the implementation of the attractive  see-saw mechanism in order to resolve
 the puzzle of the tiny neutrino masses.
 We have seen that in the present D-brane construction, the right-handed neutrino can arise
 from an open string whose one  endpoint is attached on the $U(2)$ (i.e. $D6_b$) brane-stack and
 the other on its corresponding  mirror brane-stack obtained under the orientifold action
 $\Omega {\cal R}D6_b$. Under the charge assignment of
 Table 1 a Dirac mass term for the neutrino is not allowed at the perturbative superpotential.
 The right-handed neutrino however, is an electrically  neutral singlet, while the $U(1)_b$
 charge does not play any role in the hypercharge generator, thus any of the two possible
 neutral singlets with ${\cal Q}_{b_{\nu^c}}=\pm 2$ may appear in the spectrum. Had we chosen to
 keep the neutral state with ${\cal Q}_{b_{\nu^c}}=+2$, a direct tree-level Dirac mass term
 $\lambda^{\nu}_{ij}\,L_i\,\nu^c_j\, H_u$ would be possible as was discussed in the previous
 section. In the alternative
 case of  ${\cal Q}_{b_{\nu^c}}=-2$, an effective non-perturbative term through instanton effects
 can be generated  from the ${\cal E}2-D6$ intersections of figure (\ref{f3}) provided
 $I_{{\cal E}2b}=-2$, so that the $U(1)_b$ charge excess $\sum_j{\cal Q}_{b_{j}}=-4$ is canceled
 by the instanton factor of the non-perturbative coupling.
Instanton effects can also generate Majorana masses for the singlet
right-handed neutrino which are of the form
\cite{Blumenhagen:2006xt},\cite{Ibanez:2006da},\cite{Ibanez:2007rs},\cite{Cvetic:2008hi}
\ba
M_{N,j}&=&\lambda_{N,j}\,e^{-S_{{\cal E}}}\, M_S
\ea
where  $\lambda_{N,j}=|\lambda_{N,j}|\,e^{\chi_j}$ summarizes contributions of the K\"ahler
potential and the one-loop determinant~\cite{Blumenhagen:2006xt}.

Compared to a direct tree-level perturbative mass term, the instanton generated Dirac neutrino
mass in this model is more attractive for the following two reasons. First,  there is a suppression
factor analogous to that of the Majorana right-handed neutrino mass. As a consequence, the see-saw mechanism
$m_{\nu}\propto e^{-S_{\cal E}}\frac{m_D^2}{M_S}$ can be implemented with a moderate value of the string scale.
A second observation is that in  the derivation of the neutrino mass term though instanton effects,
the corresponding disk diagrams do not involve simultaneously the matter fields
${\cal L}$ and $\nu^c$. This implies that the Dirac neutrino Yukawa coupling in the
$3\times 3$ flavor space is factorizable. To exhibit the advantage of such a simplified
 structure of the neutrino mass matrix, we assume as an example the case of almost diagonal
 charged lepton mass matrix. We further assume symmetric types of mass
matrices and  parametrise the Dirac neutrino mass matrix in terms of two real
parameters as follows. Let the vector
\ba
\vec \lambda^{\nu}=\{1,\,r\cos\theta,\,r\sin\theta\}
\ea
represent the three $\lambda_i$'s in flavor space. Then, we write the instanton induced
factorizable Dirac neutrino Yukawa coupling as follows
\ba
y_{ij}&=&e^{\imath\phi_{ij}}\lambda_i^{\nu}\lambda_j^{\nu}
\ea
where $\phi_{ij}$ are possible phases. We can write the Dirac neutrino mass matrix as follows
\ba
m_{D}^{\nu}=\left(
\begin{array}{lll}
 1 & r \cos (\theta ) & r \sin (\theta ) \\
 r \cos (\theta ) & e^{\imath\phi_1}\,r^2 \cos ^2(\theta ) &e^{\imath\phi_2} \frac{r^2}{2} \, \sin (2\theta )
   \\
 r \sin (\theta ) & e^{\imath\phi_3}\frac{r^2}{2} \, \sin (2\theta ) & e^{\imath\phi_4}r^2 \sin ^2(\theta )
\end{array}
\right)
\ea
It is easy now to pick up  choices of the parameters that reproduce the known neutrino data.
Let us for the sake of simplicity concentrate on a specific example, choosing a set of phases
$\chi_1=0$ and $\phi_i=\chi_{2,3}=\pi$. If we further assume the expansion
$r=1+\epsilon+O(\epsilon^2)$, we get
\ba
m_{\nu}^{eff.}\equiv m_{D}M_N^{-1}m_D^T&\approx&\left(
\begin{array}{lll}
 -\epsilon &  \cos (\theta ) &  \sin (\theta ) \\
  \cos (\theta ) & -\epsilon \cos ^2(\theta ) &-\frac{\epsilon}{2} \, \sin (2\theta )
   \\
  \sin (\theta ) & -\frac{\epsilon}{2} \, \sin (2\theta ) & -\epsilon \sin ^2(\theta )
\end{array}
\right)\,m_0
\ea
where $m_0$ is of the order  $e^{-S_{\cal E}}\,m_W^2/M_S$ and should be in the sub-eV range.
From the point of view of the
experimental and other observational data for neutrino masses and their mixing, this is a
very interesting example of a neutrino mass matrix. This gives bimaximal mixing and an inverted
hierarchy spectrum with two almost degenerate eigenmasses $m_{\nu_1}\sim (-1-\epsilon)m_0, m_{\nu_2}
 \sim (1-\epsilon)m_0$, and $m_{\nu_3}=0$, which, if small corrections are received from charged
 lepton mixing, could fit the neutrino data\cite{Leontaris:2005gm}.

 \section{Conclusions}

In this work we have discussed in some detail vital issues regarding the Yukawa sector in an
effective Standard Model Theory obtained from a minimal D-brane configuration~\cite{Antoniadis:2004dt}
equivalent to the gauge symmetry $U(3)\times U(2)\times U(1)$.  More precisely, we have explored the
implications  of the custodial global $U(1)$ symmetries  (indispensable in these constructions)  on the
fermion masses using sensible, phenomenologically consistent variants of the fermion and Higgs spectra.

In the desirable and elegant case of a minimal spectrum consisting just of the MSSM chiral matter,
the right-handed neutrino and the one doublet Higgs pair, several Yukawa couplings are absent in the
perturbative superpotential. It was found that the suggested mechanism based on string space-time
instanton contributions can generate the missing Yukawa couplings which fill in the zero entries
of the perturbative quark mass matrices. These corrections are found to be suppressed by an exponential
factor of the form $e^{-S_{\cal E}}$ where $ S_{\cal E}$ in the classical instanton action. This is
a generic characteristic with rather exciting and unprecedented implications on the low energy phenomenology of these constructions. Thus in a realistic D-brane Standard Model analogue one could envisage to interpret  the fermion mass spectrum  through  these  instanton effects,  attributing
the  observed  mass hierarchy to the aforementioned exponential suppression of the
non-perturbative couplings.
Furthermore, a  rather interesting Dirac neutrino mass matrix texture  is predicted and heavy Majorana
masses are also induced  from similar non-perturbative effects. As an example, we have presented a simple
parametrization of this specific texture  which easily reconciles the neutrino experimental data.

\newpage
\begin{center}
 {\bf Acknowledgements}
 \end{center}

 This work is partially supported by the European Research and Training Network MRTPN-CT-2006 035863-1
 (UniverseNet). The author would like to thank  I. Antoniadis  for useful discussions and comments.


\newpage




\begin{thebibliography}{99}


\bibitem{Antoniadis:2000ena}
  I.~Antoniadis, E.~Kiritsis and T.~N.~Tomaras,
  ``A D-brane alternative to unification,''
  Phys.\ Lett.\  B {\bf 486} (2000) 186
  [arXiv:hep-ph/0004214].
\bibitem{Aldazabal:2000sa}
  G.~Aldazabal, L.~E.~Ibanez, F.~Quevedo and A.~M.~Uranga,
  ``D-branes at singularities: A bottom-up approach to the string  embedding of
  the standard model,''
  JHEP {\bf 0008} (2000) 002
  [arXiv:hep-th/0005067].
\bibitem{Ibanez:2001nd}
  L.~E.~Ibanez, F.~Marchesano and R.~Rabadan,
  ``Getting just the standard model at intersecting branes,''
  JHEP {\bf 0111} (2001) 002
  [arXiv:hep-th/0105155].
\bibitem{Cvetic:2002wh}
  M.~Cvetic, P.~Langacker and G.~Shiu,
  ``A three-family standard-like orientifold model: Yukawa couplings and
  hierarchy,''
  Nucl.\ Phys.\  B {\bf 642} (2002) 139
  [arXiv:hep-th/0206115].
\bibitem{Kokorelis:2002qi}
  C.~Kokorelis,
  ``Exact standard model structures from intersecting D5-branes,''
  Nucl.\ Phys.\  B {\bf 677} (2004) 115
  [arXiv:hep-th/0207234].

  \bibitem{Antoniadis:2004dt}
  I.~Antoniadis and S.~Dimopoulos,
  ``Splitting supersymmetry in string theory,''
  Nucl.\ Phys.\  B {\bf 715} (2005) 120
  [arXiv:hep-th/0411032].

\bibitem{Chen:2005mj}
  C.~M.~Chen, T.~Li and D.~V.~Nanopoulos,
  ``Standard-Like Model Building on Type II Orientifolds,''
  Nucl.\ Phys.\  B {\bf 732} (2006) 224
  [arXiv:hep-th/0509059].

\bibitem{Gioutsos:2006fv}
  D.~V.~Gioutsos, G.~K.~Leontaris and A.~Psallidas,
  ``D-brane standard model variants and split supersymmetry: Unification  and
  fermion mass predictions,''
  Phys.\ Rev.\  D {\bf 74} (2006) 075007
  [arXiv:hep-ph/0605187].

  \bibitem{Leontaris:2007ej}
  G.~K.~Leontaris, N.~D.~Tracas, N.~D.~Vlachos and O.~Korakianitis,
  ``Towards a realistic Standard Model from D-brane configurations,''
  Phys.\ Rev.\  D {\bf 76} (2007) 115009
  [arXiv:0707.3724 [hep-ph]].

\bibitem{Blumenhagen:2005mu}
  R.~Blumenhagen, M.~Cvetic, P.~Langacker and G.~Shiu,
  ``Toward realistic intersecting D-brane models,''
  Ann.\ Rev.\ Nucl.\ Part.\ Sci.\  {\bf 55} (2005) 71
  [arXiv:hep-th/0502005].

\bibitem{Anastasopoulos:2006da}
  P.~Anastasopoulos, T.~P.~T.~Dijkstra, E.~Kiritsis and A.~N.~Schellekens,
  ``Orientifolds, hypercharge embeddings and the standard model,''
  Nucl.\ Phys.\  B {\bf 759} (2006) 83
  [arXiv:hep-th/0605226].

  \bibitem{Blumenhagen:2006xt}
  R.~Blumenhagen, M.~Cvetic and T.~Weigand,
  ``Spacetime instanton corrections in 4D string vacua - the seesaw mechanism
  for D-brane models,''
  Nucl.\ Phys.\  B {\bf 771} (2007) 113
  [arXiv:hep-th/0609191].

  \bibitem{Ibanez:2006da}
  L.~E.~Ibanez and A.~M.~Uranga,
  ``Neutrino Majorana masses from string theory instanton effects,''
  JHEP {\bf 0703} (2007) 052
  [arXiv:hep-th/0609213].

\bibitem{Florea:2006si}
  B.~Florea, S.~Kachru, J.~McGreevy and N.~Saulina,
  ``Stringy instantons and quiver gauge theories,''
  JHEP {\bf 0705} (2007) 024
  [arXiv:hep-th/0610003].

  \bibitem{Abel:2006yk}
  S.~A.~Abel and M.~D.~Goodsell,
  ``Realistic Yukawa couplings through instantons in intersecting brane
  worlds,''
  JHEP {\bf 0710} (2007) 034
  [arXiv:hep-th/0612110].


  \bibitem{Antoniadis:2007jq}
  I.~Antoniadis, A.~Kumar and B.~Panda,
  ``Supersymmetric SU(5) GUT with Stabilized Moduli,''
  Nucl.\ Phys.\  B {\bf 795} (2008) 69
  [arXiv:0709.2799 [hep-th]].

\bibitem{Leontaris:2008mm}
  G.~K.~Leontaris,
  ``A $U(3)_C \times U(3)_L \times U(3)_R$ gauge symmetry from intersecting D-branes,''
  Int.\ J.\ Mod.\ Phys.\  A {\bf 23}, 2055 (2008)
  [arXiv:0802.4301 [hep-ph]].
  \\
  G.~K.~Leontaris and J.~Rizos,
  ``A D-brane inspired $U(3)_C \times U(3)_L \times U(3)_R$ model,''
  Phys.\ Lett.\  B {\bf 632}, 710 (2006)
  [arXiv:hep-ph/0510230].

\bibitem{Pati:1974yy}
  J.~C.~Pati and A.~Salam,
  ``Lepton Number As The Fourth Color,''
  Phys.\ Rev.\  D {\bf 10}, 275 (1974)
  [Erratum-ibid.\  D {\bf 11}, 703 (1975)].

\bibitem{Leontaris:2000hh}
  G.~K.~Leontaris and J.~Rizos,
  ``A Pati-Salam model from branes,''
  Phys.\ Lett.\  B {\bf 510}, 295 (2001)
  [arXiv:hep-ph/0012255].

  \bibitem{Adriani:2008zr}
  O.~Adriani {\it et al.}  [PAMELA Collaboration],
  ``Observation of an anomalous positron abundance in the cosmic radiation,''
  arXiv:0810.4995 [astro-ph].

  \bibitem{Ibanez:2008my}
  L.~E.~Ibanez and R.~Richter,
  ``Stringy Instantons and Yukawa Couplings in MSSM-like Orientifold Models,''
  arXiv:0811.1583 [hep-th].

\bibitem{Ibanez:2007rs}
  L.~E.~Ibanez, A.~N.~Schellekens and A.~M.~Uranga,
  ``Instanton Induced Neutrino Majorana Masses in CFT Orientifolds with
  MSSM-like spectra,''
  JHEP {\bf 0706} (2007) 011
  [arXiv:0704.1079 [hep-th]].

  \bibitem{Cvetic:2008hi}
  M.~Cvetic and P.~Langacker,
  ``D-Instanton Generated Dirac Neutrino Masses,''
  Phys.\ Rev.\  D {\bf 78} (2008) 066012
  [arXiv:0803.2876 [hep-th]].

  \bibitem{Leontaris:2005gm}
  G.~K.~Leontaris, A.~Psallidas and N.~D.~Vlachos,
  ``Inverted neutrino mass hierarchies from U(1) symmetries,''
  Int.\ J.\ Mod.\ Phys.\  A {\bf 21} (2006) 5187
  [arXiv:hep-ph/0511327].


\end{thebibliography}
\end{document}